\documentclass[12pt,preprint]{aastex}

\newcommand{\pad}{\partial}

\newcommand{\beq}{\begin{equation}}
\newcommand{\eeq}{\end{equation}}
\newcommand{\beqn}{\begin{eqnarray}}
\newcommand{\eeqn}{\end{eqnarray}}
\newcommand{\lppr}{\stackrel{<}{\scriptstyle \sim}}
\newcommand{\gppr}{\stackrel{>}{\scriptstyle \sim}}

\slugcomment{ApJ 632 (2005), L21}
\shorttitle{Particle acceleration in GRB jets}
\shortauthors{}

\begin{document}

\title{Particle Acceleration in Gamma-Ray Burst Jets}
\author{Frank M. Rieger and Peter Duffy}
\affil{UCD School of Mathematical Sciences, University College Dublin,
       Dublin 4, Ireland}
\email{frank.rieger@ucd.ie; peter.duffy@ucd.ie}

\begin{abstract}
Gradual shear acceleration of energetic particles in gamma-ray burst 
(GRB) jets is considered. Special emphasis is given to the analysis 
of universal structured jets, and characteristic acceleration timescales 
are determined for a power-law and a Gaussian evolution of the bulk 
flow Lorentz factor $\gamma_b$ with angle $\phi$ from the jet axis. 
The results suggest that local power-law particle distributions may be 
generated and that higher energy particles are generally concentrated 
closer to the jet axis. Taking several constraints into account we show 
that efficient electron acceleration in gradual shear flows, with maximum 
particle energy successively decreasing with time, may be possible on 
scales larger than $r \sim 10^{15}$ cm, provided the jet magnetic field 
becomes sufficiently weak and/or decreases rapidly enough with distance, 
while efficient acceleration of protons to ultra-high energies $> 10^{20}$ 
eV may be possible under a wide range of conditions.
\end{abstract}

\section{Introduction}
There is mounting evidence today that gamma-ray bursts (GRBs) are associated 
with collimated relativistic outflows or jets \citep[][]{rho99,kul99,gre03}. 
It is still a matter of ongoing research, however, what types of internal 
structures are actualized within these jets. Within the fireball framework, 
for example, two different kinds of jet models seem to be compatible with the 
observations, i.e., (i) the uniform "top-hat" jet model \citep[][]{rho99,
fra01,lam04}, in which all the hydrodynamical quantities (e.g., Lorentz factor, 
energy density) are essentially the same within some well-defined opening angle 
$\phi_j$ around the jet axis but sharply drop outside of $\phi_j$, and (ii) 
the universal structured ("power-law" or "Gaussian") jet model \citep[][]{ros02,
zha02,kum03,zha04}, in which the hydrodynamical quantities are rather smooth 
functions of the angle $\phi$ from the jet axis.\\
Here we analyze the implications of such jet structures for the acceleration 
of energetic particles. We focus on shear acceleration as a promising mechanism 
for converting the kinetic energy of the flow into nonthermal particles and 
radiation. Such a mechanism has previously been successfully applied to the 
relativistic jets in Active Galactic Nuclei \citep[][]{ost98,ost00,rie04,rie05a,
rie05b}.
Shear acceleration is based on the simple fact that particles may gain energy 
by scattering off (small-scale) magnetic field irregularities with different 
local velocities due to being systematically embedded in a collisionless shear 
flow \citep[cf.][for a recent review]{rie05b}. In the case of nonrelativistic 
shear acceleration it is straightforward to show that local power-law particle 
momentum distributions $f(p) \propto p^{-\,(3 + \alpha)}$ can be generated, 
assuming a momentum-dependent mean scattering time of the form $\tau \propto 
p^{\alpha}$ with $\alpha>0$ \citep[cf.][]{berk81,rie05a}. Hence, for $\tau$ 
scaling with the gyro-radius, i.e., $\tau \propto p$, this results in a 
power-law particle number density $n(p)\propto p^2\,f(p) \propto p^{-2}$, and 
thus a synchrotron emissivity $j_{\nu}\propto \nu^{-1/2}$.

\section{Shear acceleration in structured GRB-type jets}
\begin{figure}
\plottwo{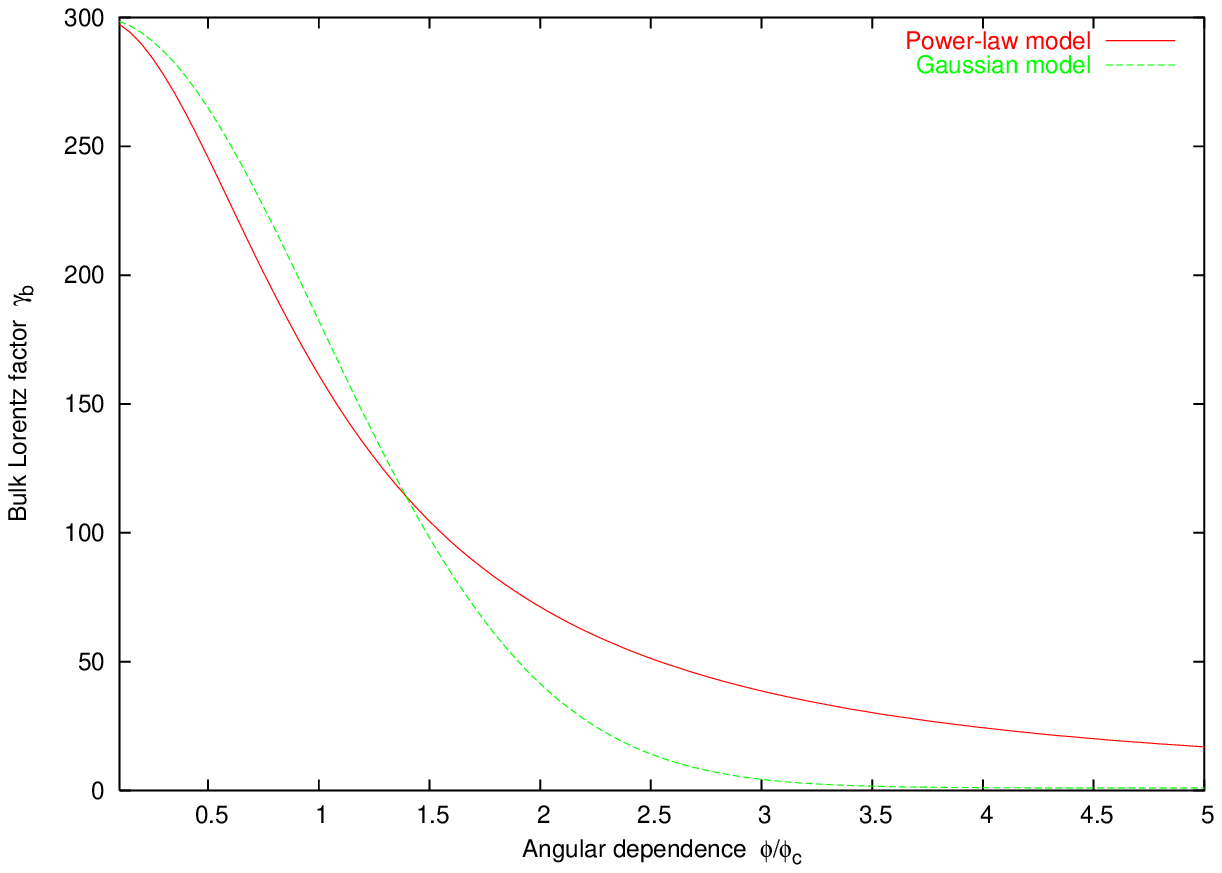}{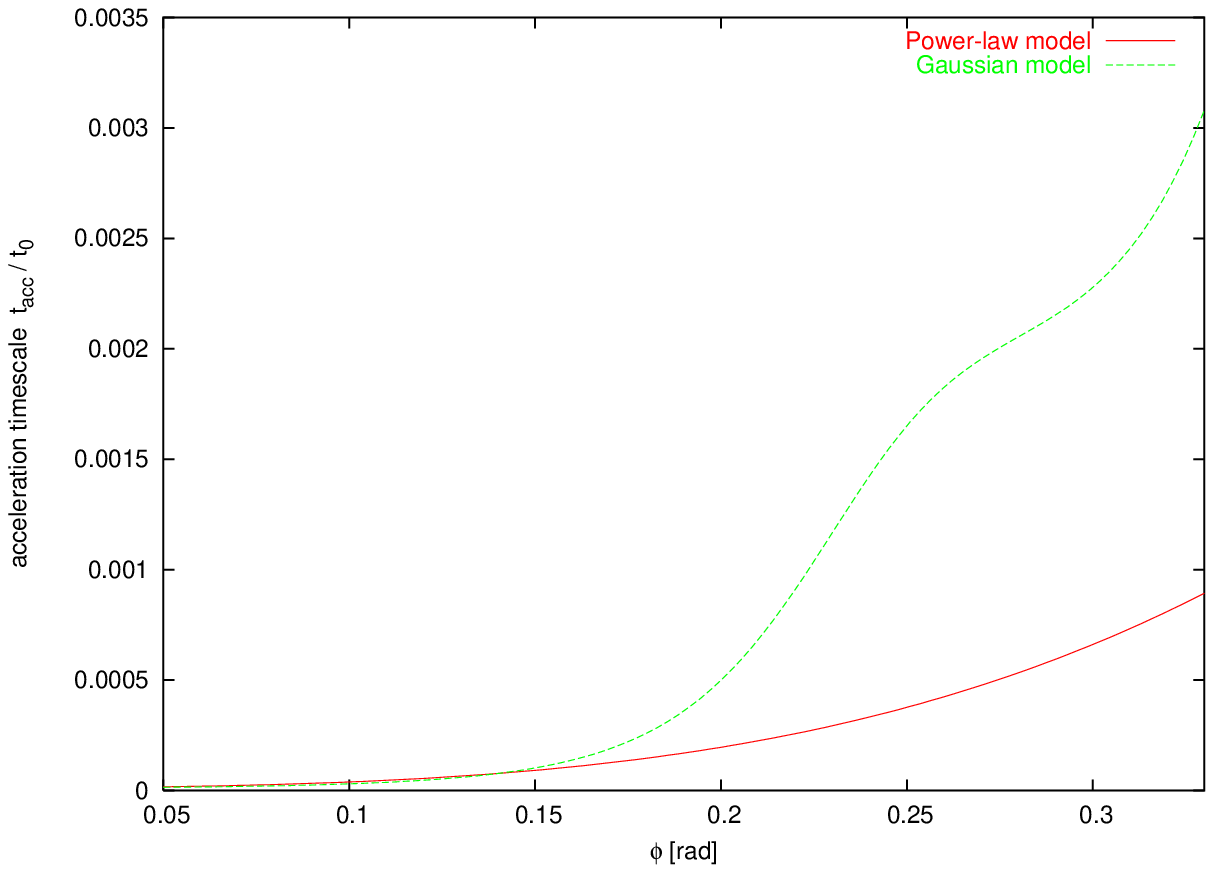}
\caption{Left: Illustration of the evolution of the bulk Lorentz factor $\gamma_b$
 with angle $\phi$ for a power-law ($b=1.8$) and Gaussian model, respectively, 
 using $\gamma_{b0}=300$. Right: Associated timescales for shear acceleration,
 calculated for $\phi_c=0.1$ [rad].\label{fig1}}
\end{figure}

Let us consider an idealized radial, relativistic flow profile, which in four 
vector notation is given by \citep[cf.][for the more general case]{rie05c}
\beq
   u^{\alpha}= \gamma_b\,\left(1,v_r(\phi)/c,0,0\right)\,,
\eeq where $\alpha=0,1,2,3$, and $\phi$ denotes the polar angle in spherical
coordinates, and $\gamma_b \equiv \gamma_b(\phi)=[1-v_r(\phi)^2/c^2]^{-1/2}$ 
is the bulk Lorentz factor of the flow. In the comoving frame, the related 
gradual shear acceleration coefficient can be cast into the form \citep[e.g.,]
[eq.~3.27]{web89}
\beq
   < \dot{p}\,'\,> = \frac{1}{p\,'^{\,2}}\frac{\pad }{\pad p\,'}
                \left(p\,'^{\,4}\,\tau'\,\Gamma\right)
\eeq where $p'$ denotes the comoving particle momentum, $\tau' \simeq \lambda'
/c$ is the mean scattering time and $\Gamma$ is the relativistic shear coefficient. 
We are interested in the strong scattering limit (corresponding to $\omega_g\,'
\,\tau' \ll 1$, with $\omega_g'$ being the relativistic gyro-frequency measured in 
the comoving jet frame, i.e., assuming a sufficiently weak longitudinal mean 
magnetic field and the presence of strong turbulence, so that collisions are 
efficient enough to restore isotropy), where the shear coefficient is given by 
$\Gamma =(c^2/30)\,\sigma_{\alpha \beta}\,\sigma^{\alpha \beta}$\citep[see][eq.~3.34]
{web89} and $\sigma_{\alpha \beta}$, with $\alpha,\beta=0,1,2,3$, is the usual 
covariant fluid shear tensor \citep[cf.][for more details]{web89,rie04,rie05c}. 
Using spherical coordinates and the velocity profile above, it can be shown 
that the relativistic shear coefficient becomes \citep[cf.][]{rie05c} 
\beq\label{shear-coeff}
  \Gamma = \frac{4}{45}\,\gamma_b^2 \left[\frac{v_r^2}{r^2}+\frac{3}{4\,r^2}\,
           \gamma_b^2\,\left(\frac{\pad v_r}{\pad \phi}\right)^2\right]\,,
\eeq which in the nonrelativistic limit ($\gamma_b \rightarrow 1$) reduces 
to the (nonrelativistic) viscous transfer coefficient derived by \citet{ear88} 
(their eq.~7) when the latter is expressed in spherical coordinates and the
corresponding velocity profile $\vec{v} = v_r(\phi)\,\vec{e}_r$ is applied. 
The (comoving) timescale $t_{\rm acc} \simeq < p' >/< \dot{p}' >$ for the 
shear flow acceleration of particles then becomes 
\beq\label{tshear}
    t_{\rm acc}(r,\phi) \simeq \frac{45}{4 (4 + \alpha)}\,
               \frac{c}{\lambda'}\,
                \frac{r^2}{\gamma_b^2\,\left[v_r^2 + 0.75\,\gamma_b^2\,   
                (\pad v_r/\pad \phi)^2\right]}\,,
\eeq where $r$ is the radial coordinate measured in the cosmological rest 
frame (i.e., the supernova, collapsar or merger rest frame), and where a power-law
dependence $\tau'=\lambda'/c \propto p'^{\alpha}$ has been assumed. As the jet 
flow is diverging, $t_{\rm acc}$ obviously increases with $r$ to the square.\\
In order to investigate the acceleration potential of structured GRB jets, two 
applications appear particularly interesting \citep[e.g.,][]{zha02,kum03,zha04}: 
a power-law model, in which $\gamma_b$ is a power-law function of $\phi$ 
outside a core of opening angle $\phi_c$, i.e., $\gamma_b(\phi) = 1 + (\gamma_{b0} 
-1)(1+[\phi/\phi_c]^2)^{-b/2}$, with $1.5 < b \lppr 2$ \citep[cf.][]{zha02}, 
and a Gaussian model with $\gamma_b(\phi) = 1 + (\gamma_{b0} -1) \exp[-\phi^2/
2\phi_c^2]$, where $\gamma_{b0}$ denotes the Lorentz factor at the jet axis, and
typically $\phi_c = 0.1$ rad \citep[][]{zha04}. The shear acceleration timescale 
(with the time and space coordinates measured in the cosmological rest frame and 
particle Lorentz factor measured in the comoving jet frame) then becomes 
\beq\label{tshear1}
  t_{\rm acc}(r,\phi) = \frac{45}{4 (4 + \alpha)}\,
      \frac{r^2}{c\,\lambda'\,\gamma_b(\phi)^2}\,\times 
      \left\{ \begin{array}{rl}
      \left(\frac{v_r^2}{c^2}+\frac{3}{4}\,\frac{\left(\gamma_b(\phi)-1\right)}
      {\gamma_b(\phi)^2 \left(\gamma_b(\phi)+1\right)}\frac{b^2}{(1+\phi^2/\phi_c^2)^2}
      \,\frac{\phi^2}{\phi_c^4}\right)^{-1} & {\rm power-law} \\
      \left(\frac{v_r^2}{c^2}+\frac{3}{4}\,\frac{\left(\gamma_b(\phi)-1\right)}
      {\gamma_b^2(\phi)\left(\gamma_b(\phi)+1\right)}\frac{\phi^2}{\phi_c^4}\right)^{-1} 
      & {\rm Gaussian\,model}
      \end{array}\,. \right. 
\eeq Note that in contrast to nonrelativistic parallel shock acceleration, 
$t_{\rm acc} \propto 1/\lambda$ as the probability of a particle sampling a higher
shear, and thus a more energetic scattering event increases with $\lambda$.
The general evolution of $t_{\rm acc}/t_0$ as a function of $\phi$, with $t_0 = 45\,r^2/[4
(4+\alpha)\,\lambda'\, c]$, is illustrated in Fig.~\ref{fig1}, indicating that a power-law 
model becomes somewhat more favourable than a Gaussian model. Obviously, the shear 
acceleration timescale in structured jets generally increases with $\phi$ due to the 
decrease in $\gamma_b$, suggesting different maximum attainable energies in the local 
comoving frame. Hence, at constant $r$ the higher energy particles and thus the higher 
energy emission are expected to be naturally concentrated closer to the jet axis, i.e., 
toward smaller $\phi$ (provided the magnetic field is not very inhomogeneous across 
the jet), enforcing the effect that observers looking at the same GRB from different 
directions will see different gamma-ray light curves.

In general several constraints need to be satisfied in order to allow for 
efficient particle acceleration in GRB jets, e.g., the particle acceleration process 
may be limited by (i) radiative synchrotron losses, (ii) the transversal confinement 
size, (iii) particle escape via cross-field diffusion or (iv) the duration of the jet 
expansion. The first constraint implies that the acceleration timescale has to be smaller 
than the comoving synchrotron cooling timescale, which, for particles with Lorentz 
factors $\gamma'$ and isotropic pitch angle distribution, is given by
\beq\label{tcool}
  t_{\rm cool} =\frac{9\,m^3\,c^5}{4\,e^4}\frac{1}{\gamma'\,B'^2}
                 \nonumber
               = 4.8\cdot 10^{12} \,\left(\frac{1}{\gamma'}\right)\,
                  \left(\frac{m}{m_p}\right)^3
                   \left(\frac{10^3\;{\rm G}}{B'}\right)^{2}\;[{\rm s}]\,,
\eeq with $m_p$ being the proton mass. According to eq.~(\ref{tshear1}) the minimum  
shear acceleration timescale is 
\beq\label{tmin}
 t_{\rm acc} \sim \frac{5}{2}\,\frac{r^2}{c\,\lambda'\,\gamma_{b}^2}\,,
\eeq so that for a gyro-dependent particle mean free path $\lambda' = \xi\, r_g'$, 
with $\xi < 1$, (as motivated by the observationally implied particle spectra with 
momentum index $\sim -2$) the ratio of shear to cooling timescale becomes independent 
of the particle Lorentz factor. Note that eq.~(\ref{tmin}) is also the timescale for a 
uniform flow with $v_r$ independent of $\phi$ (cf. eq.~[\ref{tshear}])! For a given 
magnetic field strength the acceleration process may thus work efficiently as long as 
the velocity shear remains sufficiently high, i.e., as long as the radial coordinate 
satisfies
\beq
 r < 4 \cdot 10^{15}\, \xi^{1/2}\left(\frac{m}{m_p}\right)^2 
         \left(\frac{ 10^3\;{\rm G}}{B'}\right)^{3/2} 
          \left(\frac{\gamma_{b0}}{300}\right)\;\,{\rm [cm]}\,,
\eeq suggesting that, in contrast to the acceleration of protons, efficient electron 
acceleration is suppressed in the presence of high magnetic fields. It seems very likely, 
however, that due to the expansion of the wind, the comoving magnetic field $B'$ will 
depend inversely on $r$, i.e., $B' \propto 1/r^{\beta}$ with $\beta > 0$. In order to 
study possible implications in more details, we may thus consider the following simple 
parametrization for the commoving jet magnetic field strength, $B'= 1000\,b_0\,(10^{13}
\,{\rm cm}/r)^{\beta}$ G. For $\beta=1$ and $b_0=30$ this expression corresponds 
to the lower limit required by \citet{wax95} in order to allow for efficient (second
order) Fermi-type acceleration of protons to ultra-high energies during expansion of 
the wind (cf. his eq. [4a]). While it can be shown then that proton acceleration is 
nearly unconstrained by radiative synchrotron losses, electron acceleration can only 
work efficiently if the field decays rapidly enough and becomes comparatively weak, 
e.g., for $\beta=2$ efficient electron acceleration may occur for $r \geq 10^{15} 
\xi^{-1/4}\,b_0^{3/4}(300/\gamma_{b0})^{1/2}$ cm.\\
The second constraint requires the particle mean free path $\lambda'$ to be 
smaller than the transversal width $R \sim r\,\phi_j$ of the jet (with $\phi_j > 
\phi_c$ being the jet opening angle in the cosmological rest frame). For $\lambda' = 
\xi\, r_g'$, $\xi < 1$ and $r_g'=\gamma' m\, c^2/(e\,B')$, this implies an upper 
limit for the maximum possible (comoving) particle Lorentz factor given by
\beq
\gamma_{\rm max}' \sim 10^{9} \left(\frac{1}{\xi}\right) \left(\frac{m_p}
            {m}\right) \left(\frac{B'}{10^3\;{\rm G}}\right) \left(\frac{r}
            {10^{13}{\rm cm}}\right)\left(\frac{\phi_j}{0.3\,{\rm rad}}\right)\,.
\eeq Note again, that for $B'=1000\,b_0\,(10^{13}\,{\rm cm}/r)^{\beta}$ G, this 
translates into maximum particle energies of $E \sim 2.8\cdot 10^{20}\,\xi^{-1}\,
b_0\,(10^{13}\,{\rm cm}/r)^{\beta-1}\,(\phi_j/0.3\,{\rm rad})\,(\gamma_{b}/300)$ 
eV in the observer frame. For a simple linear scaling, i.e., $\beta=1$, the maximum 
energy would become independent of $r$, while for $\beta>1$ the higher energy 
particles would originate on smaller scales.\\
The third constraint requires that the timescale for (cross-field) diffusion in the 
comoving frame $t_{\rm cf} \sim R^2/\kappa'$, with $\kappa' \sim c\,\lambda'/3$, 
is larger than $t_{\rm acc}$, i.e., $t_{\rm acc} \leq 3 \,(r^2\phi_j^2)
/(c\,\lambda')$. Comparison with eq.~(\ref{tshear1}) reveals that $t_{\rm acc}$ 
scales with $r$ and $\lambda'$ in the same way as $t_{\rm cf}$, indicating that 
the third constraint may be easily satisfied as long as $\gamma_b(\phi)$ is larger 
than a few, cf. Fig.~(\ref{fig1}).\\
Finally, the fourth constraint requires $t_{\rm acc}$ to be smaller than 
the time $t_w' =\Delta'/c$ needed to transverse the overall radial (comoving) 
width $\Delta'$ of the flow, which translates into a lower limit for the required 
particle Lorentz factors. This width is essentially determined by the activity 
of the inner engine driving the GRB and, when specified in the 
observer frame, should be at least of order of the observed GRB duration $t_{\Delta}$ 
\citep[cf.][]{pir05}, i.e., $\Delta = t_{\Delta}\,c=\Delta'/\gamma_b$. Allowing
for expansion effects on larger scales with $\Delta' \gppr r/\gamma_b$ \citep[cf.]
[]{mes93}, we have $t_w' = {\rm max}\{\gamma_b\,t_{\Delta}, r/[\gamma_b\,c]\}$.
Using Eq.~(\ref{tmin}) one thus finds
\beq\label{constraint4}
\gamma_{\rm min}' \simeq 10^4 \,\xi^{-1} \left(\frac{m_p}
                  {m}\right)\left(\frac{B'}{10^3\;{\rm G}}\right)
                  \left(\frac{300}{\gamma_b}\right)^3\left(\frac{r}{10^{13}\,
                  {\rm cm}}\right)^2 \times {\rm min}\left\{\left( \frac{10\,{\rm sec}}
                  {t_{\Delta}}\right),\chi(r)\right\} \,,
\eeq where $\chi(r)= 2.7\cdot 10^3\,(\gamma_b/300)^4\,(10^{13}\,{\rm cm}/r)$. For the 
chosen magnetic field parametrization $B'=1000\,b_0\,(10^{13}\,{\rm cm}/r)^{\beta}$ G, 
this results in $\gamma_{\rm min}' \simeq 10^4\, \xi^{-1}\,(m_p/m)\,\,b_0
\,(r/10^{13}\,{\rm cm})^{2-\beta}$ $(300/\gamma_{b})^3 \times{\rm min}\{(10\,{\rm sec}/
t_{\Delta}), \chi(r)\}$, indicating that for $\beta=2$ the leading term becomes 
independent of $r$, which may be favourable for efficient electron acceleration. 
A more restrictive condition, roughly corresponding to $\gamma_{\rm min}'$
as implied by $\chi(r)$ in eq.~(\ref{constraint4}), is usually associated with 
the requirement to overcome adiabatic losses.
Accordingly, shear acceleration can only act efficiently on particles that were 
sufficiently accelerated through other processes. In the GRB context it is possible 
that such particles can be provided, for example, by the mechanism responsible for 
the prompt burst of emission (e.g., shock acceleration).

In summary, if the jet magnetic field is sufficiently weak (e.g., $b_0 \sim 1$) 
and/or decays rapidly enough (say, e.g., with $\beta = 2$), both the first and
the fourth constraint may be satisfied even for electrons, so that the maximum 
energy is essentially determined by the third constraint. Particles may then, for 
example, be accelerated efficiently at distances larger than $r \sim 10^{12}\,
(m_p/m)$ cm (cf. constraint [i]), with possible maximum energies, measured in the 
observer frame, of less than $\sim 10^{18}$ eV for electrons and $\sim 10^{21}$ eV 
for protons, and successively (i.e., linearly for $\beta=2$) decreasing with time. 
For a high (e.g., $b_0 \sim 10$) and slowly decaying (say, e.g., $\beta \leq 1$) 
jet magnetic field, on the other hand, efficient shear acceleration of electrons is 
virtually excluded, while protons again may well be accelerated up to energies 
$>10^{20}$ eV.

\section{Comparison with shock acceleration} 
\begin{figure}
\epsscale{.60}
\plotone{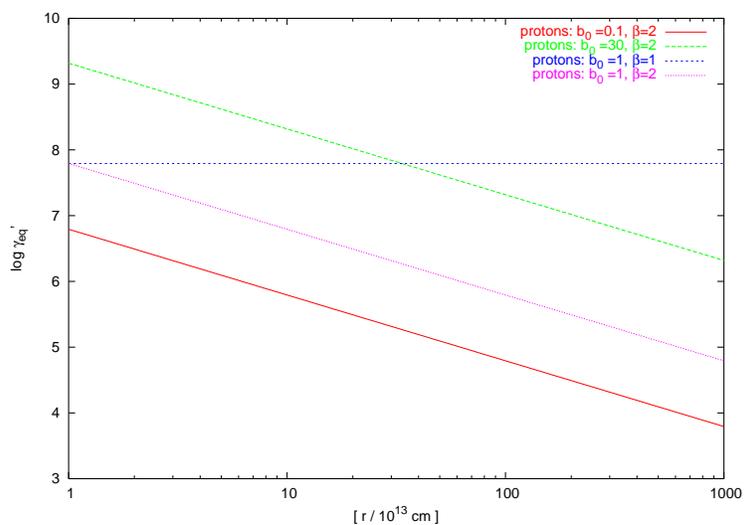}
\caption{Evolution of the critical comoving proton Lorentz factor $\gamma_{\rm 
eq}'$ as a function of the radial coordinate, shown for $\gamma_b =300$, $\xi\,\eta 
=1$, $\beta_s =0.9$, and different jet magnetic field configurations. Above 
$\gamma_{\rm eq}'$ the shear acceleration timescale becomes smaller than the shock 
acceleration timescale and proton acceleration by shear becomes more efficient 
than shock acceleration.\label{fig2}}
\end{figure}

It is widely believed that shock-accelerated electrons are responsible for the 
observed prompt GRB and afterglow emission via synchrotron radiation processes 
\citep[e.g.,][]{pir05}. In particular, diffusive electron acceleration at midly 
relativistic internal shocks (with Lorentz factor $\Gamma_s \sim$ a few), caused 
by velocity variations in the relativistic outflow is usually thought to be 
behind the powerful burst of $\gamma$-rays \citep[][]{ree94}. In the case of
nonrelativistic shocks Fermi acceleration leads to power-law particle spectra 
$N(\gamma) \propto \gamma^{-s}$, which are only dependent on the shock compression 
ratio $\rho=u_1/u_2$ (where $1 < \rho \leq 4$), i.e., $s=(\rho+2)/(\rho-1)$, so 
that for strong shocks ($\rho=4$) the famous $s=2$ result is obtained \citep[e.g.,]
[]{dru83,kir99}. In general, the acceleration timescale for diffusive shock 
acceleration depends on both the upstream and downstream residence times. For 
an unmodified nonrelativistic shock, one thus obtains \citep[e.g.,][]{dru83,jok87}
\beq 
  t_{\rm acc}=\frac{3}{u_1-u_2}\left(\frac{\kappa_1}{u_1}+\frac{\kappa_2}{u_2}\right)
             =\frac{3\,\rho}{(\rho-1)}\frac{\left(\kappa_1+\rho\,\kappa_2\right)}
                {u_s^2}\,,
\eeq with $u_1$ and $u_2$ the upstream and downstream flow velocities measured in 
the shock frame, respectively, $\kappa_1$ and $\kappa_2$ the upstream and downstream 
(spatial) diffusion coefficients, and $u_s$ the shock speed as measured in the upstream 
frame. If the acceleration process operates at nearly the Bohm limit [i.e., $\kappa 
\simeq \lambda\,c/3 = \eta\, r_g\,c/3$, with $\eta \simeq O(1)$], one finds
\beq\label{tacc2}
 t_{\rm acc} \gppr 6\, \eta\,r_g\,\frac{c}{u_s^2}\,,
\eeq assuming $\kappa_1 \simeq \kappa_2$. When $u_s\sim c$ this is comparable to 
the Larmor time which is a result also predicted by the simulations of \citet{lem03}.
It has been often suggested that the nonrelativistic limit can also be approximately
applied to midly relativistic shocks in GRBs, proposing, for example, that efficient 
proton acceleration to ultra-high energies might be possible \citep[e.g.,][]{wax04}. 
Equating the acceleration timescale (eq.~[\ref{tacc2}]), replacing $r_g$ by $r_g'$ and 
$u_s$ by $u_s'$ as the (comoving) internal shock speed, with the cooling timescale 
eq.~(\ref{tcool}), using $B'=1000\,b_0\,(10^{13}\,{\rm cm}/r)^{\beta}$ G and $u_s'
=\beta_s\,c$ with $\beta_s < 1$, gives
\beq\label{c_shock1}
 \gamma_{\rm max}' \simeq 2.5\cdot 10^9\,\eta^{-1/2}\,b_0^{-1/2}
                    \left(\frac{\beta_s}{0.9}\right)\,\left(\frac{m}{m_p}\right)\,
                    \left(\frac{r}{10^{13}\,{\rm cm}}\right)^{\beta/2}\,.
\eeq Note that magnetic field amplification in the vicinity of a shock may yield a 
field value $b_0$ well above the usual (background) flow magnetic field, thus leading 
to a somewhat more restrictive condition. In addition, the acceleration timescale 
also has to be smaller than the comoving time $\sim r/{c\,\gamma_b}$ needed to 
transverse the width of a shell, which gives
\beq\label{c_shock2}
\gamma_{\rm max}' \lppr 1.5 \cdot 10^6\,\eta^{-1}\,b_0\, \left(\frac{\beta_s}{0.9}
                  \right)^2\,\left(\frac{m_p}{m}\right)\,\left(\frac{300}{\gamma_b}
                  \right)\,\left(\frac{10^{13}\,{\rm cm}}{r}\right)^{\beta-1}\,.
\eeq 
A more detailed comparison of shear with shock acceleration may perhaps be reached 
by analyzing a critical particle Lorentz factor $\gamma_{\rm eq}'\propto b_0\,
r^{1-\beta}\,(m_p/m)$, defined by equating eq.~(\ref{tmin}) with eq.~(\ref{tacc2}), 
above which shear acceleration may become more efficient than shock acceleration.
Several applications with respect to the acceleration of protons are illustrated in 
Fig.~(\ref{fig2}), suggesting the possibility that under a reasonable range of 
conditions shear acceleration in GRB jets may become more relevant for the production 
of UHE cosmic rays than shock-type acceleration processes. Concerning electron 
acceleration in weak magnetic fields (e.g., $b_0 \leq 1$, $\beta=2$, $r \gppr 
10^{15}$ cm), on the other hand, shock-type processes would allow for a much 
quicker particle energization (but would also be more severely limited by 
eq.~[\ref{c_shock1}] and eq.~[\ref{c_shock2}]) and shear effects would only 
become dominant above $\gamma' \sim 10^8\,(10^{15}\,{\rm cm}/r)$.

\section{Conclusions}
Using an idealized model we have analyzed the possible role of shear acceleration
for the energization of particles in relativistic GRB-type jets. Our results suggest
that efficient electron acceleration on scales $r\gppr 10^{15}$ cm, with maximum 
energy decreasing with distance, may be possible in the presence of weak magnetic 
fields, assuming that high energy seed particles are provided by the mechanism 
responsible for the prompt burst of emission. This may result in a weak and long 
duration component in the GRB emission. Protons, on the other hand, may reach UHE 
energies $>10^{20}$ eV under a broad range of conditions.

\acknowledgments
We are grateful to R.~Blandford for discussions and the referee for very useful 
comments. FMR acknowledges support through a Marie-Curie Individual and a 
Cosmogrid Fellowship.

\clearpage

\clearpage

\end{document}